\documentclass[a4paper,twocolumn]{Gaia2004} 
\usepackage{times}      
\usepackage{epsfig}     
\usepackage{natbib}     
\title{Simulation of the Clock Framework of Gaia}

\author[1,2]{J. Casta\~neda}
\author[1,2]{J.P. Gordo}
\author[1,2]{J. Portell}
\author[1,2]{E. Garc\'ia-Berro}
\author[1,3]{X. Luri}
\affil[1]{Institut d'Estudis Espacials de Catalunya, 
          c/Gran Capit\`a 2-4, 08034 Barcelona, Spain}
\affil[2]{Departament de F\'isica Aplicada, 
          Universitat Polit\`ecnica de Catalunya, 
          Avda. Canal Ol\'impic s/n, 
          08860 Castelldefels, Spain}
\affil[3]{Departament d'Astronomia i Meteorologia, 
          Universitat  de Barcelona, 
          c/ Mart\'{\i} i Franqu\`es 1, 
          08028 Barcelona, Spain}

\bibpunct{(}{)}{;}{a}{}{,}  

\begin{document}

\keywords{clocks, simulation, Gaia, rubidium, Allan}

\maketitle

\begin{abstract}

Gaia  will  perform  astrometric  measurements with  an  unprecedented
resolution. Consequently, the electronics of the Astro instrument must
time   tag   every   measurement   with   a   precision   of   a   few
nanoseconds.  Hence, it requires  a high  stability clock  signal, for
which  a Rb-type  spacecraft  master clock  has  been baselined.   The
distribution  of its signal  and the  generation of  clock subproducts
must maintain  these high accuracy requirements.  We  have developed a
software application  to simulate generic clock  frameworks.  The most
critical clock structures for Gaia  have also been identified, and its
master clock has been parameterised.

\end{abstract}

\section{Introduction}

In this work  we present a powerful and  highly configurable simulator
of generic clock frameworks.   This software tool, although focused on
Gaia, has been developed as much parameterized as possible, thus being
easily  adaptable not  only to  the current  design of  Gaia  (and its
possible  modifications), but also  to almost  any space  mission. Its
objective  is to  simulate the  real performance  of an  atomic master
clock and  its several sub-products.  It generates  a realistic signal
for  the master  clock, including its  typical noises.   It also
simulates its  distribution and  the generation of  clock sub-products
from this master signal,  using devices such as frequency multipliers,
frequency dividers  or transmission lines. Finally the resulting clock
outputs  taken from  several nodes  of the  framework are displayed, both
graphically and numerically. These  outputs can be used for validating
several design  issues of Gaia,  such as the synchronization  lines of
the  payload, the timing  and codification  schemes, the  time tagging
accuracy of the CCD measurements, or the typical deviation between two
daily  calibrations.   The parameters  of  the  master  clock and  the
framework  devices, as  well  as the  clock  framework structure,  are
entered using XML files and can be graphically verified.

\section{Basic concepts}

\subsection{Specification and analysis of clock signals}

The  main parameters  used  to  characterise a  clock  signal are  the
following:

\begin{itemize}
\item Accuracy:  The degree of conformity  of a measured  value to its
      definition or with respect to a reference.
\item Frequency  drift:  The   linear  (first-order)  component  of  a
      systematic change in frequency of an oscillator with time. Drift
      is due to ageing and to changes in the environment.
\item Frequency offset: The  frequency difference between the measured
      value and the nominal (pre-determined) value.
\item Precision:  The  degree of  mutual agreement  among a  series of
      individual measurements.
\item Resolution:   The  degree   to  which  the  measurement  can  be
      determined.
\item Frequency  stability: The statistical estimate  of the frequency
      fluctuations of a signal within a given time interval.
\end{itemize}

The  frequency stability  cannot be  measured by  using  the classical
definition  of  standard  deviation.   Instead,  a slightly different  
version of  the standard deviation  that basically behaves as a filter 
to  many noise components is  often  used  . This analysis tool, the
so-called Allan variance (or Allan deviation), can be used to identify
noise  sources, types of  oscillators, and  noises of  the measurement
system.  There  are two  versions of the  Allan deviation,  Eqs.~\ref{eq:allanstd} and
~\ref{eq:allanmod} provide their general definition.

\begin{eqnarray}
\sigma_{\rm Std} (\tau) &=& \sqrt{\frac{1}{2} \frac{1}{n} \sum_{n=0}^{N-1} 
{\Delta^2 \Phi_{\tau}^2(n)} } \label{eq:allanstd} \\
\sigma_{\rm Mod} (\tau) &=& \sqrt{\frac{1}{2} \frac{1}{n} \sum_{n=0}^{N-1} 
{\Delta^2 \langle\Phi\rangle_{\tau}^2(n)} }
\label{eq:allanmod}
\end{eqnarray}

\begin{figure}[t]
\begin{center}
\leavevmode
\centerline{\epsfig{file=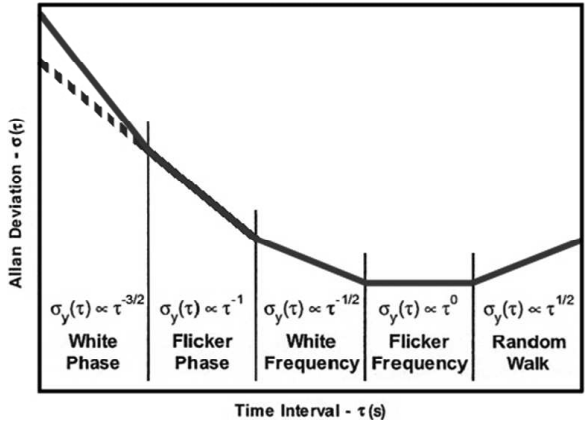,width=8.0cm}}
\end{center}
\caption{Standard and modified Allan deviations.}
\label{fig:AllanDevGrl}
\end{figure}

In these expressions  $\tau$ is the measurement interval  of the clock
phase,  $\Delta\Phi$ is  the  first finite  difference  of the  phase,
$\Delta^2\Phi$      is     the      second      finite     difference,
$\Delta^2\langle\Phi\rangle$ is an  average of the frequency stability
and  $n$  is  the  number  of  samples  averaged  when  computing  the
variance. As shown  in Figure~\ref{fig:AllanDevGrl} the standard Allan
variance (dashed  line) cannot  distinguish between white  and flicker
phase noise,  whereas the modified  version (solid line) results  in a
different slope for each one.

\subsection{Rubidium Clocks}

\begin{figure}[t]
\begin{center}
\leavevmode
\centerline{\epsfig{file=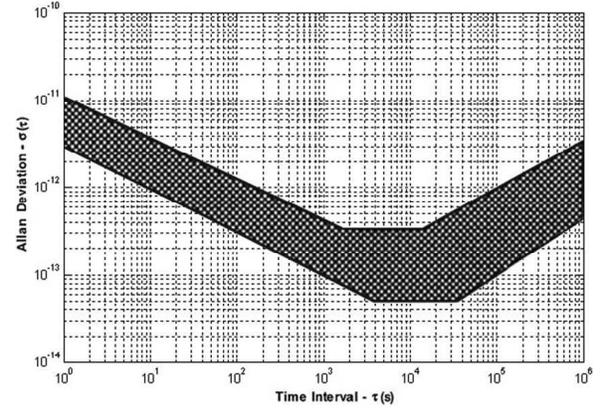,width=8.0cm}}
\end{center}
\caption{ Typical Allan deviation profile of a Rb clock.}
\label{fig:AllanDevRb}
\end{figure}

The use of a  Rb atomic clock  as the  spacecraft master clock of Gaia 
is currently envisaged.  For a proper study of this master clock it is 
necessary to simulate the response  of these types  of oscillators with their  typical noise
sources  and  parameters. Table~\ref{tab:RbSpecs} \citep{CastaPFC} lists the typical parameters of a Rb atomic clock, while  Figure~\ref{fig:AllanDevRb} illustrates its typical Allan variances.

\begin{table}[t]
  \caption{Typical parameters for a Rb atomic clock.}
  \label{tab:RbSpecs}
  \begin{center}
    \leavevmode
        \begin{tabular}[h]{ll}
      \hline \\[-5pt]
      Parameter           & Value \\[+5pt]
      \hline \\[-5pt]
      Resonance frequency & 6~834~682~608 \\
      Output frequency    & 1 to 15 MHz \\
      Frequency stability & 5 x 10$^{-12}$ over 1 sec \\
                          & 5 x 10$^{-12}$ over 100 sec \\
                          & 1 x 10$^{-12}$ over 10~000 sec \\
      Frequency drift     & 1 x 10$^{-13}$ per day \\
                          & 3 x 10$^{-11}$ per month \\
                          & 5 x 10$^{-10}$ per year \\
      Ageing              & 3 x 10$^{-10}$ per year \\
      \hline \\
      \end{tabular}
  \end{center}
\end{table}

\subsection{Simulation of oscillators and clock devices}

The output of any generic oscillator can be modeled using 

\begin{eqnarray}
V(t) &=& [V_0 + \epsilon (t)] \sin [\Phi (t)] \nonumber \\
\Phi (t) &=& 2 \pi (\nu_0 + \Delta \nu)t + \pi D \nu_0 t^2 + 
\phi (t) + \Phi_0\nonumber
\end{eqnarray}

\noindent  where  $V_0$  is  the  nominal  amplitude  of  the  signal,
$\epsilon(t)$  is  the  amplitude  noise, $\Delta\nu$  represents  the
frequency  offset of  the  actual clock  from  the nominal  frequency,
$\nu_0$; $D$ is the  linear fractional frequency drift rate; $\phi(t)$
is the  random phase  deviation, modelling oscillator  intrinsic phase
noise sources,  and $\Phi_0$  is the initial  phase offset.  The basic
noise components which contribute to $\phi(t)$ are:

\begin{itemize}
\item White  phase,  which is  due to  broadband noise  from amplifier
      stages and components.
\item Flicker phase, which are a consequence of noisy components.
\item White  frequency,  which is  often found  when an  oscillator is
      locked to a Ce or Rb standard.
\item Flicker  frequency, caused  by the  resonator noise,  the active
      component noise, or both.
\item Random  walk,  which is  due  to environmental  factors such  as
      mechanical  shocks,  vibrations  and  temperature  fluctuations,
      which cause random shifts in frequency.
\end{itemize}

All these  noises can  be described by  power-law noises  ($S_\nu \sim
\nu^\alpha$).  For $\alpha\ge$1 these  noises present a power spectral
density  that   cannot  be  integrated,  thus   implying  an  inherent
non-stationary behavior of the underlying process. Additionally, these
noises have another  important characteristic: their scale invariance,
this  is,  the  power law  process  is  independent  of the  scale  of
observation.   They can  be simulated  using the  spectral  shaping of
white  noise  samples,  which  can  be obtained  with  the  Box-Muller
method.  This operation can  be performed  by algorithmic  and digital
signal processing techniques, such  as the Voss-McCartney Algorithm or
the Fractional Brownian Motion model.

\section{Overview of clocks in Gaia}

\begin{figure}[t]
\begin{center}
\leavevmode
\centerline{\epsfig{file=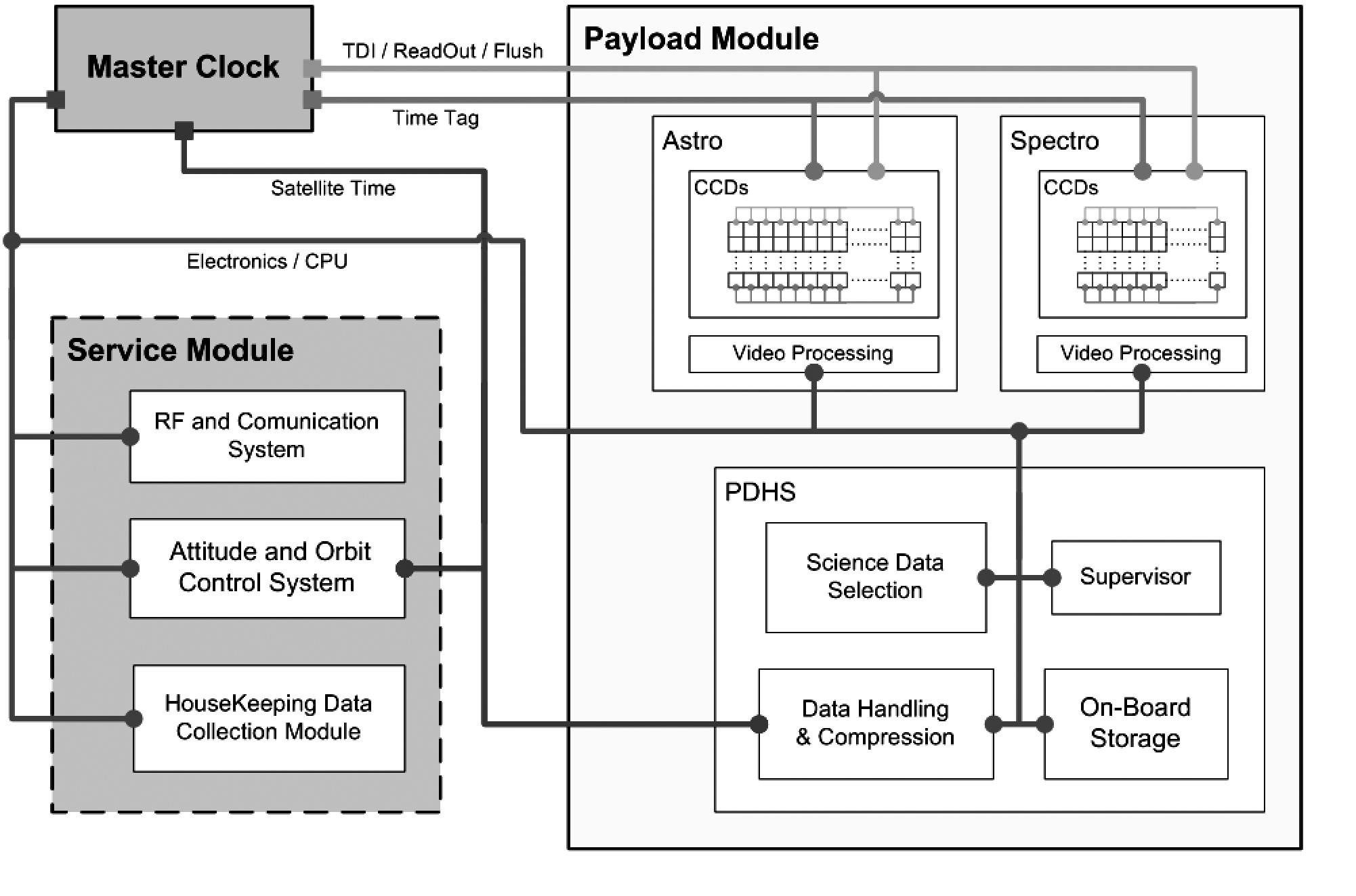,width=8.0cm}}
\end{center}
\caption{Overview of the clock distribution lines in the payload of Gaia.}
\label{fig:ClockDistrGaia}
\end{figure}

In  Gaia   there  will  coexist   several  clock  signals   which  are
interrelated.  However,  all of them  are derived from  the spacecraft
Master  Clock.  These  clock  signals  control the  operation  of  the
instruments  (Astro and  Spectro),  the data  processing pipeline  and
several specific functions  of the satellite. We focus  on the Payload
Data Handling System,  and on the effects of  non-ideal clock signals.
Table~\ref{tab:cksignals} summarizes  the most relevant  clock signals
to  analyse  and   their  approximate  operating  frequencies.   Their
distribution   in   the   payload    of   Gaia   is   illustrated   in
Figure~\ref{fig:ClockDistrGaia}.

\begin{table}[t]
  \caption{Main clock signals in the payload of Gaia.}
  \label{tab:cksignals}
  \begin{center}
    \leavevmode
    \small{
        \begin{tabular}[h]{lcc}
      \hline \\[-5pt]
      Name             & Frequency Range & Requirements \\[+5pt]
      \hline \\[-5pt]
      Electronics/CPU  & 6.4 to 100MHz   & Low\\
      Satellite time   & 1Hz             & Medium\\
      MC Ticks counter & 6.4MHz          & Medium/high \\
      CCD pixel flush  & 6.4MHz          & High\\
      TDI and readout  & 50kHz to 1MHz   & High\\
      Time tags        & 500MHz          & Critical\\
      \hline \\
      \end{tabular} }
  \end{center}
\end{table}

\section{Clock framework simulator}

We have developed a generic Clock Framework Simulator able to generate
any master clock signal  including power-law noises.  Its distribution
with transmission  lines and distribution nodes is  also simulated, as
well  as   the  generation  of  clock   sub-products  using  frequency
multipliers  and dividers.   All of  these elements  add noise  to the
signal  and, additionally,  will degrade  it, so  the final  result is
quite realistic.

\begin{figure}[t]
\begin{center}
\leavevmode
\centerline{\epsfig{file=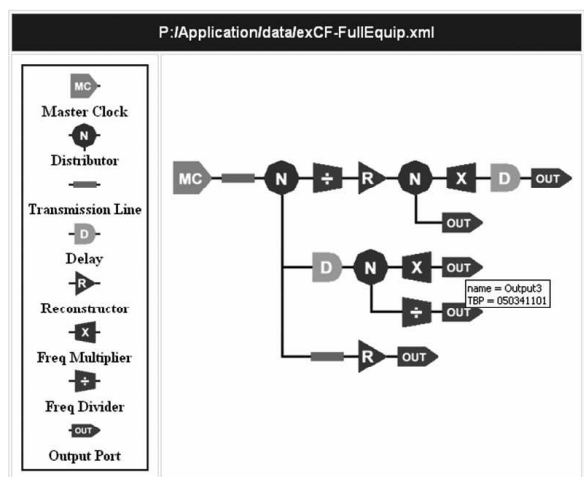,width=8.0cm}}
\end{center}
\caption{Example of a clock framework configuration as seen with our software tool.}
\label{fig:FramewEx}
\end{figure}

The  entire simulator  is configured  through a  stand-alone  XML file
where we can define not only  the structure of the clock framework but
also  the specific  parameters  of each  device.  The clock  framework
configuration  can  be  validated  and reviewed  through  a  graphical
interface  of  our application,  a  sample  of  which is  included  in
Figure~\ref{fig:FramewEx}.

\begin{figure}[ht]
\begin{center}
\leavevmode
\centerline{\epsfig{file=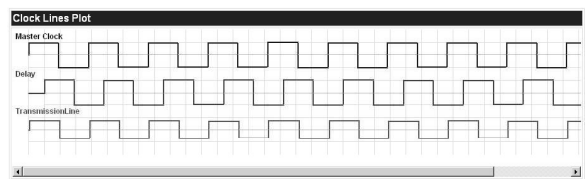,width=8.0cm}}
\end{center}
\caption{Signal visualisation tool included in our software application.}
\label{fig:SigVisTool}
\end{figure}

\begin{figure}[ht]
\begin{center}
\leavevmode
\centerline{\epsfig{file=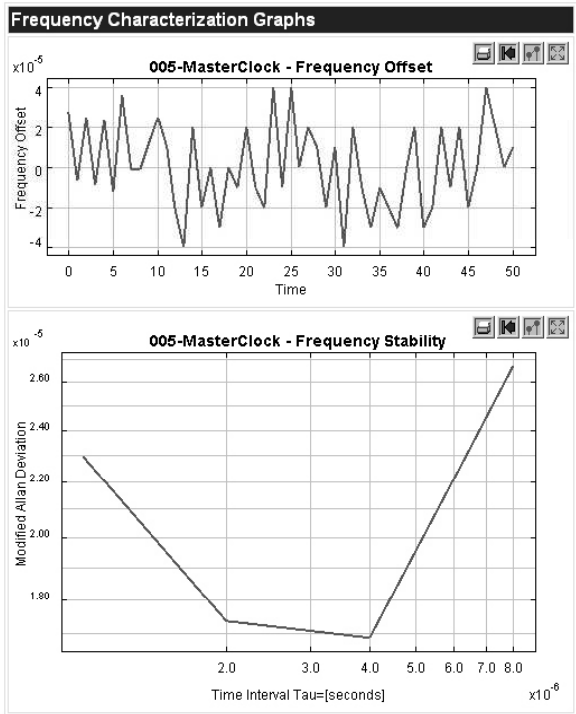,width=8.0cm}}
\end{center}
\caption{Statistical analysis tools included in our software application.}
\label{fig:StatVisTool}
\end{figure}

The user can specify the format  of the output file to be generated by
the  simulator, as  well as  different output  nodes (each  offering a
different  format if  necessary).  Time  delays and  waveforms  can be
directly verified  with the  {\tt .stt} and  {\tt .ccl}  formats, even
directly     feeding    these     data     to    other     simulators.
Figures~\ref{fig:SigVisTool}    and    \ref{fig:StatVisTool}   contain
snapshots of  the visualisation tools  developed for their  study.  On
the other hand, the simulator can directly offer the Allan deviations,
offering  an {\tt .adv}  file that  can be  plotted with  any standard
visualisation tool.

\section{Simulation of the Master Clock of Gaia}

\begin{figure}[ht]
\begin{center}
\leavevmode
\centerline{\epsfig{file=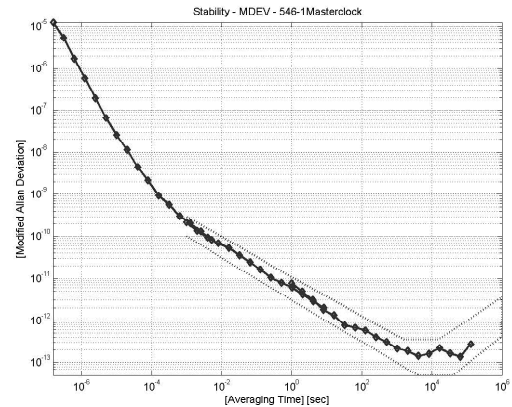,width=8.0cm}}
\end{center}
\caption{Allan Deviation of the Gaia Master Clock as obtained with our simulator.}
\label{fig:GaiaMC}
\end{figure}

Our main objectives included the simulation of a realistic Gaia Master
Clock, and  the typical deviations at $\tau$  = 1 day, 1  second and 1
Astro TDI  (736$\mu$s). For this, we  had to simulate  long periods of
time with  our tool.  The time  required to obtain  an Allan deviation
point  is proportional to  the Master  Clock frequency,  so simulation
times required to obtain the  deviation after 1 day were prohibitively
large  (up to some  years). Taking  into account  the scale-invariance
property of power-law  noises, a solution to this  problem consists in
simulating  at  lower MC  frequencies  with  equivalent noise  sources
(rescaling  their effect).   With  this approach,  we  split the  main
simulation into much  faster sub-simulations, each  with an  equivalent Master
Clock  at  different frequencies  and  thus  covering  a given  $\tau$
range.  Overlapping  their plots  we  obtain  the  final Master  Clock
characterisation  shown   in  Figure~\ref{fig:GaiaMC}  \citep{BCN010}.
From  this, we  can obtain  the  desired typical  deviations from  the
nominal frequency:

\begin{enumerate}
\item $\sigma$=2x10$^{-13}$ over 1 day (1.28$\mu$Hz deviated)
\item $\sigma$=6x10$^{-12}$ over 1 second (38.4$\mu$Hz deviated)
\item $\sigma$=3x10$^{-10}$ over 1 Astro TDI period, i.e., 736$\mu$s (1.92mHz deviated)
\end{enumerate}

\section{conclusions}

We have compiled the models  that define a typical oscillator with its
corresponding   noises,   focusing  on   the   Rb   atomic
oscillator class. Their  typical parameters, including  the Allan deviation
profile,  have  been  identified.  Several simulation  techniques  for
generating noisy clock signals have been tested, selecting the best of
them for each noise type. Finally, the most relevant clock signals and
their possible distribution in Gaia have been identified.

Additionally,  a versatile  software application  has  been developed,
which is easily configured with  XML files. This software can simulate
most of the typical clock frameworks, not only that of Gaia. Graphical
user interfaces  make possible the verification of  the designed clock
framework and  its parameters. The  simulation results, the  format of
which can  be  selected, can also  be graphically analysed.  As a
first realistic test of this  new simulation tool, the Master Clock of
Gaia has been characterised, parameterised and its Allan deviation
has been obtained.  For this,  we have determined the relation between
the  frequency scaling  of the  Master Clock  and the  scaling  of its
noises. At the end, a realistic Master Clock signal has been generated
and the typical deviations after different integration times have been
determined.

\section*{Acknowledgements}

This   work  has   been  partially   supported  by   the   MCYT  grant
AYA2002--4094--C03--01, by  the European Union FEDER funds  and by the
CIRIT.

\end{document}